\documentclass[conference]{IEEEtran}
\IEEEoverridecommandlockouts

\usepackage[T1]{fontenc}
\usepackage{cite}
\usepackage{amsmath,amssymb,amsfonts}
\usepackage{algorithmic}
\usepackage{graphicx}
\usepackage{textcomp}
\usepackage{xcolor}
\usepackage{booktabs}
\usepackage{url}
\usepackage{placeins}
\usepackage{hyperref}
\hypersetup{hidelinks, pdftitle={Cohort Effects, Not Treatment Response, Dominate Public Immunotherapy Cohorts: Evidence from Microbiome and Tumor Genomic Data}, pdfauthor={Amey Garg}, pdfkeywords={cohort effects, batch effects, microbiome, PERMANOVA, immunotherapy}}

\def\BibTeX{{\rm B\kern-.05em{\sc i\kern-.025em b}\kern-.08em
 T\kern-.1667em\lower.7ex\hbox{E}\kern-.125emX}}

\newcommand{\bd}{\texttt{batch\_detector}}

\begin{document}

\title{Cohort Effects, Not Treatment Response, Dominate Public Immunotherapy Cohorts: Evidence from Microbiome and Tumor Genomic Data%
\thanks{This is an extended version of a paper accepted to appear in the IEEE BIBM 2026 Workshop Proceedings. The shorter conference version is the version of record.}}

\author{\IEEEauthorblockN{Amey Garg}
\IEEEauthorblockA{\textit{Santa Clara High School} \\
Santa Clara, CA, USA \\
ameygarg52@gmail.com}
}

\maketitle

\begin{abstract}
Researchers often pool public gut microbiome datasets from immunotherapy studies to get around small sample sizes, assuming that adding cohorts adds statistical power faster than it adds differences between studies. We test that assumption. We processed four public shotgun metagenomic cohorts of patients receiving immune checkpoint blockade through one pipeline and pooled them step by step, from 39 to 283 patients. Because the PERMANOVA effect size $R^2$ is above zero even when there is no real effect, we compared every result with its exact chance expectation. Measured as raw variance explained, cohort membership accounted for 8 to 27 times more variance than treatment response, although much of that ratio comes from sample size alone. At every pooling step, treatment response explained no more variance than chance (within 0.2 percentage points) and was never significant, while cohort membership explained 6.0 to 9.5 percentage points beyond chance ($p = 0.001$ each time). Classification stayed at chance across three model types, none of six leak-free batch-correction strategies recovered signal, and no genus reached significance in a random-effects meta-analysis. The largest single cohort ($n=165$) was also non-significant, although simulation shows it had at least 90\% power to detect an effect at its own 95\% upper bound; within it, recruiting site alone explained 7.0 percentage points beyond chance. In three tumor genomic cohorts ($n=223$), a small but real response effect sat next to a cohort effect about eighteen times larger, so the problem is not specific to microbiome data. We release \bd{}, an open tool that reports chance-corrected cohort and response effects before pooling, and test it on an independent colorectal cancer meta-analysis.
\end{abstract}


\section{Introduction}

Immune checkpoint inhibitors, drugs that block the PD-1 and PD-L1 pathways so the immune system can attack tumors, have changed the treatment of metastatic melanoma and several other cancers. However, many patients do not benefit, and they still face immune-related side effects and high treatment costs. Predicting who will respond before treatment starts is therefore a central problem in precision oncology.

Most prediction efforts have focused on two biological systems. The first is the tumor itself: markers such as tumor mutational burden and PD-L1 expression are established but imperfect predictors \cite{b_chan}. The second is the gut microbiome. Several influential studies reported that the bacteria living in a patient's gut are associated with response to checkpoint blockade \cite{b_sivan,b_gopal,b_matson,b_routy}. Later meta-analyses reported microbial signatures that were partly shared across cohorts \cite{b_limeta,b_olekh} but also strongly dependent on cohort and treatment context \cite{b_lee,b_zhang}. These findings suggested a simple computational strategy: combine the public sequencing data from these studies, train a classifier on the pooled data, and overcome the small sample size that limits each individual study.

This strategy depends on an assumption that is rarely tested: that pooling independent cohorts adds statistical power faster than it adds differences between studies. Individual immunotherapy microbiome studies are small, usually fewer than fifty labeled patients. They also differ in sequencing machine, sample preparation, DNA extraction, location, tumor type, and even how they define a responder. We use the term \emph{cohort effect} for all of these systematic differences between studies, both technical and biological. We use \emph{batch correction} for the standard methods designed to remove them. These effects are well documented in microbiome research \cite{b_duvallet,b_wirbel,b_nearing}. What has not been measured for this heavily studied problem is how big the cohort effect is compared with the response effect, whether standard corrections are enough, and what data would be needed for pooling to work.

In this work, we treat the act of pooling itself as the thing being studied. We assembled four public gut microbiome cohorts of checkpoint blockade patients, processed them through one identical pipeline, and measured how much variance was explained by response and by cohort as we added cohorts one at a time, from 39 patients to 283. One statistical point turned out to be central. PERMANOVA, the standard test for whether groups of samples differ in overall composition, reports an effect size $R^2$. This $R^2$ is not zero when there is no real effect. For a factor with $g$ groups, its expected value by chance is $(g-1)/(n-1)$. This is the same bias that the adjusted $R^2$ corrects in ordinary ANOVA, and it is well known there, but it is rarely applied when PERMANOVA $R^2$ values are compared across cohort sizes or across factors. This means raw $R^2$ shrinks as $n$ grows even when nothing is there, and a factor with more groups gets a larger share just by chance. Comparing $R^2$ across sample sizes, or between a four-group cohort factor and a two-group response factor, only makes sense after removing this chance level. We do that throughout this paper. We also repeat the analysis on tumor genomic data from three melanoma studies to test whether the pattern is specific to microbiome data.

The main result is that treatment response explains no more variance than chance at any cohort size, while cohort membership explains several percentage points beyond chance at every step. This is a null result for prediction, and we report it as one. Careful negative and methodological results have an established place in microbiome research \cite{b_duvallet,b_nearing}, where they have corrected overly optimistic conclusions from small single studies.

Our contributions are:

\begin{itemize}
\item A step-by-step, chance-corrected measurement of cohort and response variance from 39 to 283 patients. Response stays at its chance level at every step and within each cohort alone, while cohort stays 6.0 to 9.5 percentage points above chance. Both results hold at equal sequencing depth, across zero-handling methods and taxonomic levels, and in a joint model.
\item Evidence that the null holds at three levels: overall community composition, classification with three model types, and the effects of individual bacterial genera.
\item A leak-free comparison of six batch-correction strategies, none of which recovers signal, plus a calibrated simulation showing how large a signal must be before pooling helps.
\item Evidence that the null is not caused by too few patients: the largest single cohort had at least 90\% simulated power to detect an effect at its own 95\% upper bound, and still found nothing.
\item A tumor genomic replication, where a real but small response effect is outweighed by a cohort effect about eighteen times larger. This shows the problem comes from study design, not from microbiome data specifically.
\item A practical point for any multi-cohort PERMANOVA analysis: the standard chance correction for $R^2$, routine in ANOVA, is needed whenever PERMANOVA $R^2$ is compared across sample sizes or factors, because raw $R^2$ drops, and raw $R^2$ ratios grow, as $n$ increases.
\item \bd{}, an open command-line tool that reports chance-corrected cohort and response effects and a pooling recommendation for any multi-cohort omics dataset, tested on a colorectal cancer meta-analysis.
\end{itemize}

\section{Materials and Methods}

\subsection{Cohort Assembly}

We assembled four public gut microbiome cohorts of patients receiving immune checkpoint blockade (Table~\ref{tab:cohorts}). All four used paired-end Illumina shotgun metagenomic sequencing. However, they used three different instrument models and were collected at different sites under different protocols.

Cohort C1 \cite{b_frankel} is complete. Its BioProject contains 44 runs, which correspond to 39 unique patients plus five technical replicates, so no more same-protocol samples exist. Cohort C2 \cite{b_peng} contains gastrointestinal cancer patients rather than melanoma patients, all receiving anti-PD-1 or anti-PD-L1 therapy. We kept it on purpose. Combining whatever public immunotherapy microbiome data exists, across tumor types, is exactly the practice this paper evaluates, so the differences it adds are part of what we are measuring. Cohort C4 \cite{b_lee} was recruited at five sites in three countries.

The curation we used for response labels \cite{b_orlet} covers 624 samples from 11 public studies. We processed four of them (C1--C4, 283 samples). The other seven (341 samples; Gopalakrishnan 2018, Gunjur 2024, Heshiki 2020, Liu 2022 (a microbiome study, unrelated to the Liu 2019 tumor cohort), McCulloch 2022, Spencer 2021, and Tsakmaklis 2023) were not processed in this study. All seven have public raw reads, but they differ from C1--C4 in ways that would need separate handling. Three include treatments outside our anti-PD-1/PD-L1 scope (combined nivolumab and ipilimumab in Gunjur 2024; chemotherapy with or without immunotherapy across eight cancers in Heshiki 2020; anti-PD-1 plus anti-CTLA-4 in 13 of 29 patients in Tsakmaklis 2023), and McCulloch 2022 treated 14 patients with pembrolizumab plus pegylated interferon. The runs the curation lists for Gopalakrishnan 2018 are stored under the Spencer 2021 project, so these two may share patients. Together, the curation's subsets of these studies come to about 1~TB of raw reads. Our conclusions therefore apply to these four cohorts, and extending the analysis to the remaining studies is the most direct next step. Our C3 and C4 each contain one more sample than the curation lists (39 versus 38, and 165 versus 164); the extra two (SRR6000943 in C3, labeled responder, and ERR10290768 in C4, labeled non-responder) take their labels from the original studies' label files. Dropping both ($n=281$) left the results essentially unchanged: response excess $+0.03$ points ($p=0.305$) and cohort excess 9.4 points ($p=0.001$).

Response labels for all four cohorts came from this single published curation. It defines a responder as complete response, partial response, or stable disease at six months, following RECIST 1.1. Using one curation avoids adding our own labeling inconsistencies. However, its definition is broader than the complete-or-partial-response definition used in the original C1 and C3 papers, and we did not re-check individual labels against those papers. Tumor type, site, and response definition are therefore all part of the cohort effect, along with sequencing platform.

\begin{table}[!htbp]
\caption{Gut Microbiome Cohorts}
\begin{center}
\begin{tabular}{llllc}
\toprule
\textbf{ID} & \textbf{Study} & \textbf{Platform} & $\boldsymbol{n}$ & \textbf{R/NR} \\
\midrule
C1 & Frankel 2017 \cite{b_frankel} & HiSeq 2000 & 39 & 24/15 \\
C2 & Peng 2020 \cite{b_peng}$^{\mathrm{a}}$ & NovaSeq 6000 & 40 & 25/15 \\
C3 & Matson 2018 \cite{b_matson} & NextSeq 500 & 39 & 15/24 \\
C4 & Lee 2022 \cite{b_lee}$^{\mathrm{b}}$ & NovaSeq 6000 & 165 & 64/101 \\
\midrule
\multicolumn{3}{l}{\textbf{Total}} & \textbf{283} & \textbf{128/155} \\
\bottomrule
\multicolumn{5}{l}{$^{\mathrm{a}}$Gastrointestinal cancer; all others metastatic melanoma.} \\
\multicolumn{5}{l}{$^{\mathrm{b}}$Five recruiting sites across three countries.} \\
\multicolumn{5}{l}{Labels harmonized from a single curation \cite{b_orlet}; see text.} \\
\multicolumn{5}{l}{Accessions: C1 PRJNA397906; C2 PRJNA615114;} \\
\multicolumn{5}{l}{C3 PRJNA399742; C4 PRJEB43119.}
\end{tabular}
\label{tab:cohorts}
\end{center}
\end{table}

\subsection{Bioinformatics Pipeline}

All 283 samples went through one identical pipeline, so any differences between cohorts come from the data rather than from analysis choices. We quality-filtered and adapter-trimmed the raw paired-end FASTQ files with fastp \cite{b_fastp}. We then classified reads taxonomically with Kraken2 v2.17.1 \cite{b_kraken} against the k2\_standard\_08gb database (March 2023 build, hash.k2d MD5 9d57fbd9f8bcee3d646991df60ed1f66). This 8~GB database is a capped subset of the full standard database and is less sensitive. Our cohorts were sequenced to different depths (median read pairs after fastp: 41.1, 25.3, 35.7, and 20.6 million for C1--C4) and had different fractions of reads classified, excluding human (medians 52\%, 63\%, 75\%, and 55\%). Finally, we combined the Kraken2 reports into a genus-level table of clade percentages (relative abundances) and removed human reads. For a depth check, we also built a table of integer genus read counts from the same reports and rarefied every sample to the smallest total, 2,098,257 reads (Section~\ref{sec:zero}).

Microbiome counts are compositional, meaning each sample only tells us relative amounts, not absolute ones. We therefore applied a centered log-ratio (CLR) transform to each sample vector $\mathbf{x}$ of $D$ taxa,
\begin{equation}
\mathrm{clr}(\mathbf{x})_i = \ln\!\left[\frac{x_i + \epsilon}{G}\right], \quad G = \left( \prod_{j=1}^{D} (x_j + \epsilon) \right)^{1/D}
\label{eq:clr}
\end{equation}
where $G$ is the geometric mean of the shifted counts. Our main analysis used $\epsilon = 10^{-6}$. A pseudocount this small turns every zero into a large negative number, so distances can end up driven by which taxa are present or absent, and that depends on sequencing depth and database. We therefore repeated all variance analyses with $\epsilon = 0.5$, with $\epsilon = 1$, and with multiplicative zero replacement \cite{b_martin} (Section~\ref{sec:zero}). All distances use the Aitchison distance, which is the ordinary Euclidean distance after the CLR transform.

\subsection{Tumor Genomic Pipeline}

To test whether our findings are specific to microbiome data, we separately assembled three melanoma anti-PD-1 cohorts from cBioPortal \cite{b_cbio}: Riaz 2017 ($n=42$), Liu 2019 ($n=144$), and Hugo 2016 ($n=37$), for 223 patients total. The features were tumor mutational burden plus 63 binary gene-level mutation flags. These features are not compositional counts, so we used Euclidean distance on standardized features instead of Aitchison distance. This data type shares no features with the microbiome data and differs in sample type, assay, and biology.

\subsection{Variance Partitioning and Chance Correction}
\label{sec:varpart}

We used PERMANOVA \cite{b_anderson2001} with 999 permutations to measure how much variation between samples is explained by treatment response and by cohort membership. The effect size $R^2$ is the fraction of the total variation explained by a factor.

$R^2$ is above zero even when a factor has no real effect. If the group labels are shuffled at random, the expected within-group sum of squares is $SS_T (n-g)/(n-1)$ no matter how big each group is. So the expected $R^2$ under shuffling is exactly
\begin{equation}
E_0[R^2] = \frac{g-1}{n-1}.
\label{eq:null}
\end{equation}
For example, a two-group response factor at $n=39$ is expected to explain 2.6\% of the variance just by chance. A four-group cohort factor gets three times the chance share of a two-group factor at the same $n$. We checked this formula directly: at $n=118$, the average $R^2$ across 999 label shuffles was 0.00865 for response ($E_0 = 0.00855$) and 0.01707 for cohort ($E_0 = 0.01709$). For every result, we report $R^2$, its chance level $E_0[R^2]$, and the excess $\Delta R^2 = R^2 - E_0[R^2]$. The excess is about zero when a factor has no real effect. It is the familiar adjusted $R^2$, $R^2_{\mathrm{adj}} = 1-(1-R^2)(n-1)/(n-g)$, up to a constant factor: $R^2_{\mathrm{adj}} = \Delta R^2\,(n-1)/(n-g)$. We report $\Delta R^2$ because it is on the same scale as $R^2$.

Because raw ratios like this are often reported, we also report
\begin{equation}
\rho = R^2_{\mathrm{cohort}} \big/ R^2_{\mathrm{response}},
\label{eq:ratio}
\end{equation}
but we do not rely on it. When response has no effect, $R^2_{\mathrm{response}} \approx 1/(n-1)$, so $\rho \approx R^2_{\mathrm{cohort}}(n-1)$. This grows almost linearly with $n$ even if the cohort effect stays the same. Also, when neither factor has an effect, $\rho$ is expected to be about $g_{\mathrm{cohort}}-1$, not one. We therefore compared each observed $\rho$ to a permutation null in which both labels were shuffled independently, 999 times.

In our data, cohort and response are related, because the fraction of responders differs by cohort (62\% in C1, 38\% in C3). Two separate single-factor models can give the shared variance to both factors. We therefore also fit a joint model, $\text{distance} \sim \text{cohort} + \text{response}$. We report the marginal $R^2$ of each term, meaning what it explains after accounting for the other. The response term was tested by shuffling labels only within each cohort, so the cohort structure is kept intact under the null. We implemented this directly in Python using the hat-matrix method of McArdle and Anderson \cite{b_mcardle}. For the response term, this is equivalent to \texttt{adonis2(dist $\sim$ cohort + response, by="margin", permutations=how(blocks=cohort))} in the R package vegan. The cohort term was tested with unrestricted shuffling, since cohort is the blocking factor.

PERMANOVA can confuse a difference in group centers with a difference in how spread out the groups are. We therefore ran PERMDISP \cite{b_anderson}, a test for differences in spread, alongside the main PERMANOVA tests. We computed confidence intervals on $R^2$ with a parametric Dirichlet bootstrap. Naive resampling of the distance matrix creates duplicate samples with zero distance between them, which inflates the test statistic. In our data, it produced intervals that did not even contain their own point estimates. Resampling the compositional weights instead fixes this.

\subsection{Classification and Validation}

We ran the analyses step by step: $n=39$ (C1 alone), then $n=79$, $n=118$, and $n=283$ as cohorts were added, plus C4 alone. We used three classifiers: elastic net, random forest, and gradient boosted trees. Our main protocol was repeated stratified five-fold cross-validation, repeated ten times for 50 outer folds in total. Hyperparameters were tuned by an inner five-fold loop inside each outer fold, and AUC, the area under the ROC curve where 0.5 means chance, was computed within each outer test fold. We also ran leave-one-out cross-validation (LOOCV) with nested tuning as a secondary check. When there is no real signal, LOOCV tends to give AUCs below 0.5. This happens because removing one sample shifts the class balance of the training set against that sample's label \cite{b_parker,b_airola}. Below-chance LOOCV AUCs should be read with this in mind.

Every preprocessing step that learns from the data, including feature selection, batch correction, and scaling, was fit on the training fold only. In an early analysis, we applied batch correction to all the data before cross-validation and got an apparent random forest AUC of 0.995. We recognized this as leakage and discarded it. We assessed significance with permutation tests ($N=100$) that shuffled the labels and re-ran feature selection and batch correction inside every fold of every shuffled run. To keep run time reasonable, we did not re-run the inner hyperparameter search for each shuffle. Instead, we used the hyperparameters chosen most often across the 50 real outer folds. In this protocol, the pooled sets were mean-centered by cohort within each training fold; C1 and C4 alone needed no correction.

\subsection{Batch Correction and Simulation}

We compared six strategies, each applied leak-free within folds: no correction, mean centering, location-scale adjustment, quantile mapping, percentile normalization, and adding cohort as a model covariate.

To tell whether these methods failed because of the methods or because of the data, we built a simulation where the right answer is known. We generated samples in CLR space from a multivariate Gaussian, using a Ledoit-Wolf shrinkage covariance estimated from the $n=118$ data (75 most common genera). We specified effect sizes as target PERMANOVA $R^2$ values, written $R^2_{\mathrm{t}}$. We varied signal effect size ($R^2_{\mathrm{t}} \in \{0.007, 0.027, 0.05, 0.10\}$), cohort effect size ($R^2_{\mathrm{t}} \in \{0, 0.04, 0.08, 0.15\}$), samples per cohort (20, 40, 80), and number of cohorts (1 to 4). We tested six correction methods with five-fold cross-validation and ten repeats per setting. The lowest signal level, $R^2_{\mathrm{t}} = 0.007$, was calibrated to match the pooled $n=118$ response $R^2$. Since that $R^2$ is at chance, this level has essentially no injected signal, and it represents our real data.

\subsection{Meta-Analysis of Individual Genera}
\label{sec:metamethods}

For each genus present in at least 10\% of a cohort's samples, we computed Hedges' $g$, a standardized difference in mean CLR abundance between responders and non-responders, along with its standard error. We combined these across the four cohorts with a DerSimonian-Laird random-effects model \cite{b_dersimonian}, for genera present in at least two cohorts. We corrected for multiple testing with the Benjamini-Hochberg false discovery rate (FDR). We report $I^2$, the percentage of variation in estimates that comes from real differences between cohorts rather than from chance.

We also estimated the smallest pooled Hedges' $g$ this design could detect with 80\% power, using the standard fixed-effect approximation with each cohort's variance $1/n_R + 1/n_{NR}$.

\subsection{Power}
\label{sec:powermethods}

We estimated statistical power for \emph{hypothetical} effect sizes instead of for the observed effect. The observed effect is at chance, so basing a sample-size estimate on it would be circular \cite{b_hoenig}. We did the power analysis in C4, the largest single cohort, in two feature spaces: its 30 most common genera, and its full genus set (3,005 genera), which is the feature space of the C4 PERMANOVA we report. We used the Dirichlet generator to create new samples and added a calibrated CLR shift to 10 common genera, so that the expected $R^2$ equals $E_0[R^2] + \Delta R^2$ for target $\Delta R^2 \in \{0.0025, 0.005, 0.0078, 0.01\}$. Power is the fraction of 200 simulated datasets with permutation $p < 0.05$. In the full genus set we computed power at $n \in \{80, 131, 165, 250, 400\}$; in the 30-genus space we computed it at $n=165$. Separately, we bootstrapped C4's real data 500 times, using its 30 most common genera, to get a 95\% upper bound on its true response effect.

\section{The batch\_detector Tool and External Validation}
\label{sec:tool}

Any group planning a pooled study would benefit from running the analysis above first. We packaged it as \bd{}, an open command-line tool that answers one question: given several cohorts, is it safe to pool them? The name follows common usage, but the tool measures cohort effects in the broad sense defined above.

The tool takes a feature table, response labels, and cohort labels, and works with any assay. It can apply the CLR transform in \eqref{eq:clr} with a choice of zero handling, and then computes distances. For both cohort and response, it reports the observed $R^2$, the chance level from \eqref{eq:null}, the excess $\Delta R^2$, and permutation $p$-values. It fits the joint model from Section~\ref{sec:varpart}, with response shuffled only within cohort. It reports $\rho$ together with its percentile in a permutation null, so the raw ratio is never read on its own. It also reports power as a curve over hypothetical effect sizes, not as one sample size based on the observed effect. The output is a JSON report, a power table, and a figure. The tool only needs NumPy, pandas, SciPy, and Matplotlib; the joint model and multiplicative zero replacement are written directly instead of relying on R or scikit-bio. On a laptop (Apple M3 Pro, 18~GB RAM), it took 3.6 minutes on our $n=283$ dataset and 13.0 minutes on the 570-sample validation set, mostly spent on the permutation-based power grid.

The tool's recommendation is based on chance-corrected numbers. It says NO-GO if the response excess $\Delta R^2 \le 0$, if the response effect is not significant, if $\rho > 4$, or if the observed $\rho$ is at or above the 95th percentile of its permutation null. It says GO if the response excess is positive and significant and $\rho \le 0.25$, and CAUTION otherwise. The cutoffs of 0.25 and 4 are cautious rules of thumb, not values derived from the simulation. In the simulation, pooled models reliably beat the single-cohort ceiling only at the largest signal size we tested ($R^2_{\mathrm{t}} = 0.10$), and whether they did so did not depend on the ratio of signal to cohort effect: they did so when the signal was smaller than the cohort effect and when there was no cohort effect at all. So there is no single ratio at which pooling becomes safe, and the tool's cutoffs are meant to be conservative.

We ran \bd{} unchanged on a colorectal cancer meta-analysis it was not built on: four cohorts, 570 samples, 16S amplicon sequencing instead of shotgun, and a disease unrelated to immunotherapy \cite{b_duvallet}. Disease status was significant ($p=0.001$), with $R^2 = 0.0062$ against a chance level of 0.0018, an excess of 0.44 percentage points. Cohort membership explained $R^2 = 0.245$ against a chance level of 0.005, an excess of 23.9 points. The tool returned NO-GO.

Comparing this dataset with ours shows what the rule is designed to do. Both get NO-GO, but for different reasons. In our pooled microbiome data, the response effect cannot be told apart from chance, so pooling has nothing to preserve. In the colorectal data, a real signal exists. The NO-GO comes instead from $\rho = 39.5$ sitting at the 100th percentile of its permutation null: a model trained on the combined data would be dominated by a cohort effect about fifty times larger than the disease effect. Whether a signal is significant and whether pooling is safe are different questions, and a rule based only on raw ratios cannot tell them apart.

\textbf{Availability.} Source code, documentation, and the analysis scripts for this paper are available at \url{https://github.com/ames943/batch_detector} under the MIT license, archived on Zenodo at \url{https://doi.org/10.5281/zenodo.23226392}.

\section{Results}
\label{sec:results}

\subsection{Response Variance Stays at Chance at Every Cohort Size}

\begin{figure*}[!t]
\centerline{\includegraphics[width=\textwidth]{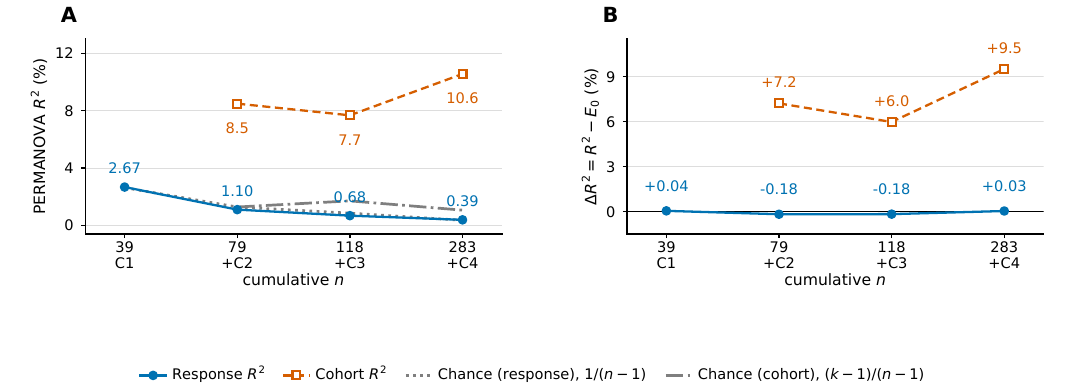}}
\caption{Response and cohort variance compared with chance as cohorts are pooled. (A) Observed PERMANOVA $R^2$ for treatment response (filled circles) and cohort membership (open squares), with each factor's chance level: $1/(n-1)$ for response (dotted) and $(k-1)/(n-1)$ for cohort (dash-dot), where $k$ is the number of cohorts pooled. Response sits on its chance curve at every $n$. (B) Excess variance $\Delta R^2 = R^2 - E_0[R^2]$. Response stays at zero while cohort stays six to ten percentage points above it. The horizontal axis is ordered, not scaled to $n$. Cohort is undefined at $n=39$, which is a single cohort.}
\label{fig:dose}
\end{figure*}

Table~\ref{tab:permanova} and Fig.~\ref{fig:dose} show the main result. Response $R^2$ was 2.67, 1.10, 0.68, and 0.39\% at $n = 39$, 79, 118, and 283. The chance levels at those sizes were 2.63, 1.28, 0.85, and 0.35\%. The excess was within 0.2 percentage points of zero at every step, and at $n=79$ and $n=118$ it was actually negative, meaning response explained less than chance. So the drop in response $R^2$ as cohorts are added is exactly the drop expected for a factor with no effect. It does not mean pooling is destroying a real signal, because there was no detectable signal in any single cohort: response excess was $+0.04$, $+0.23$, $-0.25$, and $+0.05$ percentage points in C1--C4 alone, with $p = 0.419$, 0.275, 0.635, and 0.277 (Table~\ref{tab:percohort}). The Dirichlet bootstrap interval for C1, [0.0181, 0.0295], contains both the observed value and the chance level.

Cohort membership was different. It explained 7.2, 6.0, and 9.5 percentage points beyond chance in the three pooled sets, with $p=0.001$ each time, the smallest $p$-value possible with 999 permutations. The raw ratio $\rho$ rose from 7.7 to 27.3, but most of that rise is just arithmetic. With response at chance, $\rho \approx R^2_{\mathrm{cohort}}(n-1)$, which gives 6.6, 9.0, and 29.8 from the observed cohort $R^2$ values. Against a permutation null where both labels were shuffled independently, the observed $\rho$ was at the 100th percentile at all three pooling levels, with null means of 1.06, 2.09, and 3.12. So $\rho$ is larger than random labels would produce, but its size should not be read as a biological fold-change.

\begin{table*}[!t]
\caption{Chance-Corrected PERMANOVA Across Step-by-Step Pooling (Genus Level, Aitchison Distance, 999 Permutations, $\epsilon=10^{-6}$; $E_0[R^2] = (g-1)/(n-1)$; $p=0.001$ Is the Minimum Possible; Multiply $R^2$ by 100 for \%)}
\label{tab:permanova}
\centering
\begin{tabular}{llccccccccc}
\toprule
 & & \multicolumn{4}{c}{Response} & \multicolumn{4}{c}{Cohort} & \\
\cmidrule(lr){3-6}\cmidrule(lr){7-10}
$n$ & Cohorts & $R^2$ & $E_0[R^2]$ & $\Delta R^2$ & $p$ & $R^2$ & $E_0[R^2]$ & $\Delta R^2$ & $p$ & $\rho$ \\
\midrule
39  & C1     & 0.0267 & 0.0263 & $+0.0004$ & 0.419 & --     & --     & --     & -- & -- \\
79  & C1--C2 & 0.0110 & 0.0128 & $-0.0018$ & 0.736 & 0.0849 & 0.0128 & $+0.0721$ & 0.001 & 7.7 \\
118 & C1--C3 & 0.0068 & 0.0085 & $-0.0018$ & 0.874 & 0.0768 & 0.0171 & $+0.0597$ & 0.001 & 11.3 \\
283 & C1--C4 & 0.0039 & 0.0035 & $+0.0003$ & 0.284 & 0.1056 & 0.0106 & $+0.0950$ & 0.001 & 27.3 \\
\bottomrule
\end{tabular}
\end{table*}

\begin{figure*}[!t]
\centerline{\includegraphics[width=\textwidth]{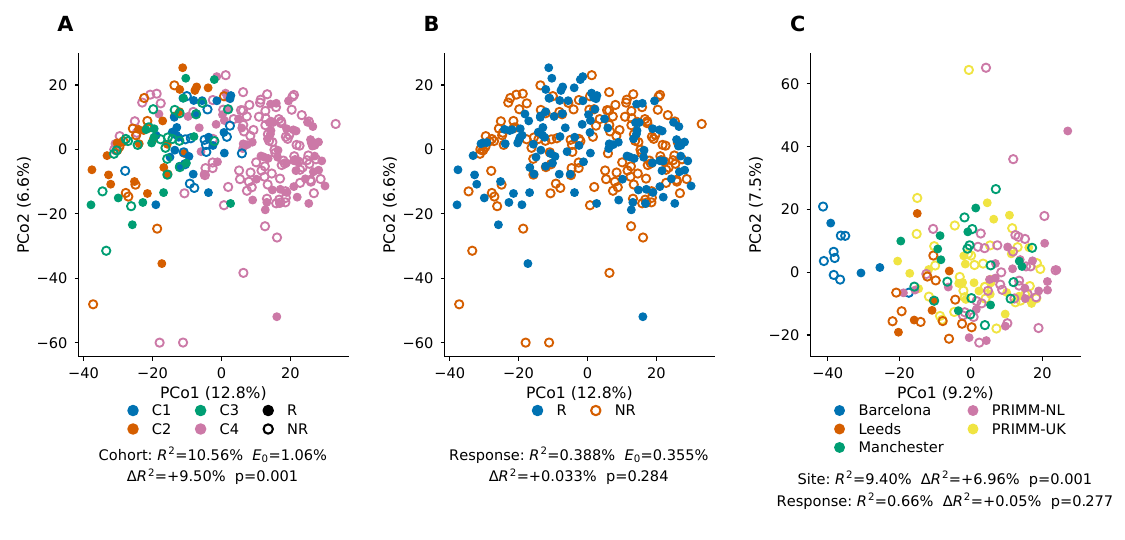}}
\caption{Principal coordinates analysis of Aitchison distances. (A) All 283 samples colored by cohort; filled markers are responders and open markers are non-responders. (B) The same coordinates colored by response only. (C) C4 alone ($n=165$) colored by recruiting site. Statistics below each panel are from Table~\ref{tab:permanova} and Section~\ref{sec:c4}. Samples group by study and by site, not by treatment response.}
\label{fig:pcoa}
\end{figure*}

Fig.~\ref{fig:pcoa} shows the same result visually. In a principal coordinates plot, a standard way to draw high-dimensional distances in two dimensions, the 283 samples group by cohort, with C4 forming its own region (A). The same plot colored by response shows no visible pattern (B).

PERMDISP found that responders and non-responders were equally spread out at every sample size ($p = 0.341$, 0.192, 0.779, 0.366), so the response results are not affected by differences in spread. Cohorts were not equally spread out at $n=79$ ($F=9.59$, $p=0.005$), $n=118$ ($F=5.63$, $p=0.006$), or $n=283$ ($p=0.001$). This means cohorts differ in both their center and their spread, which location-scale corrections are not designed to fix.

\subsection{Joint Model}

Fitting cohort and response together changes nothing (Table~\ref{tab:joint}). With response labels shuffled only within cohort, the marginal response term explained 1.11, 0.58, and 0.35\% of the variance at $n=79$, 118, and 283, with $p = 0.590$, 0.959, and 0.312. The marginal cohort term still explained 8.50, 7.58, and 10.52\% at $p=0.001$. These numbers are close to the single-factor results in Table~\ref{tab:permanova}, so the separate models were not inflating either factor.

In the two datasets where a real response effect exists, the joint model keeps it. The tumor cohorts gave a marginal response $R^2$ of 0.71\% ($p=0.025$), and the colorectal validation set gave 0.62\% ($p=0.001$). Both stayed significant after accounting for cohort.

\begin{table}[!htbp]
\caption{Joint Model, Marginal $R^2$, Response Shuffled Within Cohort}
\label{tab:joint}
\begin{center}
\begin{tabular}{lccccc}
\toprule
Dataset & $n$ & Resp. $R^2$ & $p$ & Cohort $R^2$ & $p$ \\
\midrule
Microbiome & 79  & 0.0111 & 0.590 & 0.0850 & 0.001 \\
Microbiome & 118 & 0.0058 & 0.959 & 0.0758 & 0.001 \\
Microbiome & 283 & 0.0035 & 0.312 & 0.1052 & 0.001 \\
Tumor      & 223 & 0.0071 & 0.025 & 0.1379 & 0.001 \\
CRC 16S    & 570 & 0.0062 & 0.001 & 0.2446 & 0.001 \\
\bottomrule
\end{tabular}
\end{center}
\end{table}

\subsection{Zero Handling, Sequencing Depth, and Taxonomic Level}
\label{sec:zero}

Neither the zero-handling method nor the taxonomic level changes the conclusion, although the size of the cohort effect does vary. With $\epsilon = 0.5$, $\epsilon = 1$, and multiplicative replacement, the cohort excess stayed large and significant at every pooling level, from 7.7 to 18.4 percentage points, so it is robust in sign and significance while its size varies by up to a factor of two. The response excess stayed within about 0.6 points of zero and was never significant (Table~\ref{tab:zero}). The tiny pseudocount in our main analysis is actually the cautious choice for the cohort effect, because all three alternatives make the cohort effect bigger.

Sequencing depth explains part of the cohort effect but not most of it, and does not change the response result. Genus-assigned reads per sample ranged from 2.1 to 56.2 million (quartiles 7.9, 12.8, and 19.6 million). After rarefying all 283 samples to the minimum, 2,098,257 reads, response excess was $+0.0005$ percentage points ($p=0.336$) and cohort excess was 7.0 percentage points ($p=0.001$), compared with $+0.03$ and 9.5 points without rarefaction. Rarefying instead to 5 million reads, which dropped 29 low-depth samples (all from C4, leaving $n=254$), gave a response excess of $+0.04$ points ($p=0.186$) and a cohort excess of 8.4 points ($p=0.001$). So equalizing depth removed between about a tenth and a quarter of the cohort effect, depending on the depth chosen, and response stayed at chance in both cases.

At $n=118$, response $R^2$ was at or below its chance level of 0.0085 at the phylum, genus, and species levels. The cohort excess ranged from 4.0 to 6.0 percentage points, even though the number of features ranged from 65 to 8,051 (Table~\ref{tab:taxa}). PERMDISP again found responders and non-responders equally spread at all three levels ($p = 0.969$, 0.762, 0.267), and cohorts unequally spread ($p = 0.008$, 0.002, 0.015).

\subsection{Classification Fails Across Models and Corrections}

\begin{figure*}[p]
\centerline{\includegraphics[width=\textwidth]{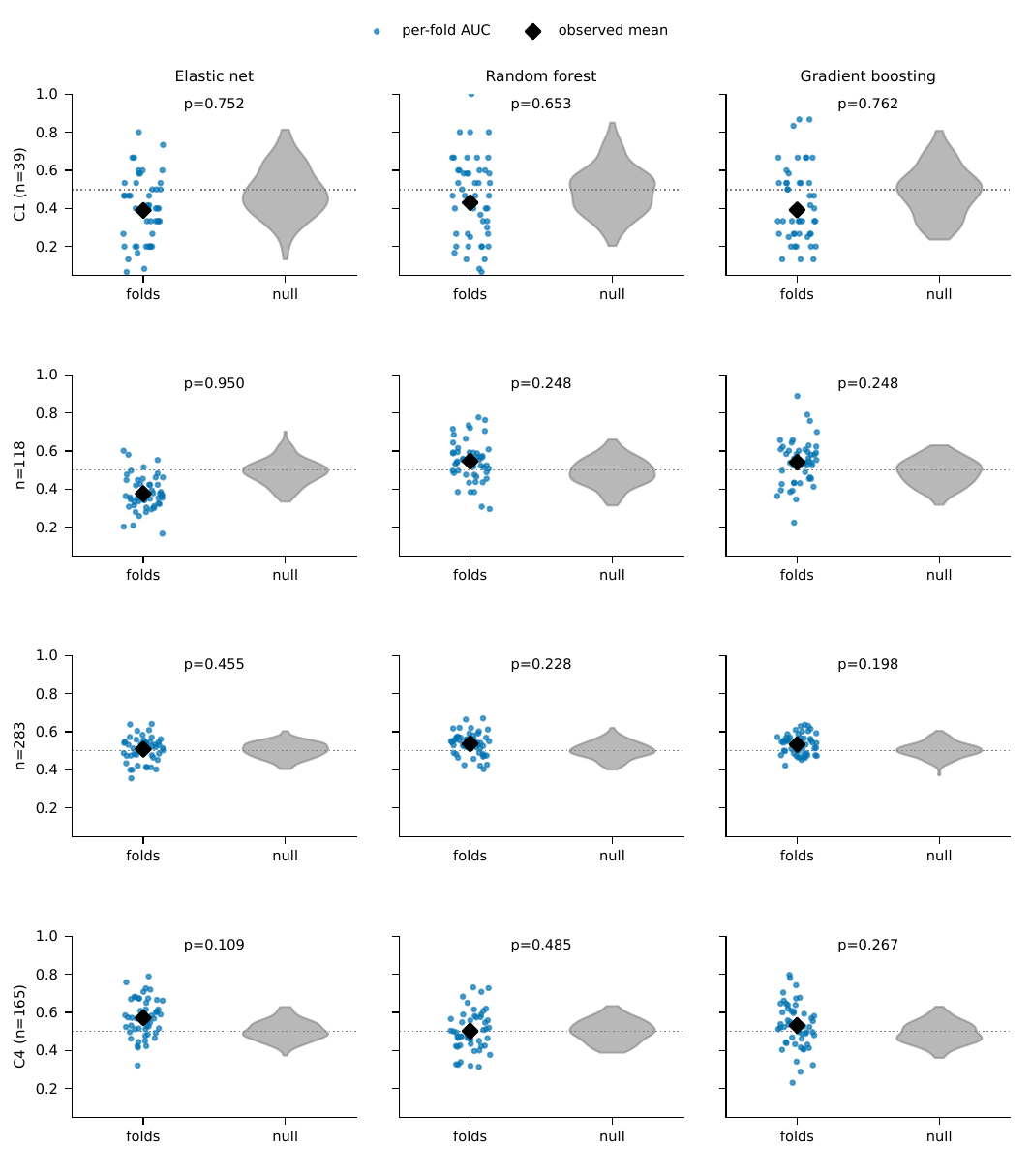}}
\caption{Classification under repeated stratified five-fold cross-validation (10 repeats, 50 outer folds). For each dataset and model, blue points are the 50 per-fold test AUCs and the black diamond is their mean. The gray violin is the permutation null distribution of the mean AUC ($N=100$). The diamond, not the spread of the points, is what should be compared with the violin. Dotted line: AUC 0.5.}
\label{fig:cv}
\end{figure*}

No model, dataset, or correction method found a predictive signal. Under the main repeated five-fold protocol, all twelve model-dataset combinations were non-significant (Table~\ref{tab:repcv}, Fig.~\ref{fig:cv}). Mean AUCs ranged from 0.376 to 0.571. The closest to significance was elastic net on C4 alone (0.571, $p=0.109$). The pooled sets reached at most 0.546 ($n=118$, random forest, $p=0.248$) and 0.537 ($n=283$, random forest, $p=0.228$).

\begin{table}[!htbp]
\caption{Repeated Stratified 5-Fold CV (10 Repeats, 50 Outer Folds; Permutation $N=100$; EN: Elastic Net, RF: Random Forest, GB: Gradient Boosting)}
\label{tab:repcv}
\begin{center}
\begin{tabular}{llccc}
\toprule
Dataset & Model & AUC ($\pm$ SD) & Perm. mean & $p$ \\
\midrule
C1 ($n=39$)   & EN & $0.390 \pm 0.180$ & 0.485 & 0.752 \\
              & RF & $0.431 \pm 0.229$ & 0.495 & 0.653 \\
              & GB & $0.392 \pm 0.215$ & 0.491 & 0.762 \\
$n=118$       & EN & $0.376 \pm 0.090$ & 0.491 & 0.950 \\
              & RF & $0.546 \pm 0.107$ & 0.488 & 0.248 \\
              & GB & $0.541 \pm 0.117$ & 0.488 & 0.248 \\
$n=283$       & EN & $0.508 \pm 0.063$ & 0.502 & 0.455 \\
              & RF & $0.537 \pm 0.060$ & 0.502 & 0.228 \\
              & GB & $0.533 \pm 0.053$ & 0.504 & 0.198 \\
C4 ($n=165$)  & EN & $0.571 \pm 0.099$ & 0.504 & 0.109 \\
              & RF & $0.502 \pm 0.104$ & 0.502 & 0.485 \\
              & GB & $0.531 \pm 0.120$ & 0.491 & 0.267 \\
\bottomrule
\multicolumn{5}{l}{SD is across 50 correlated folds, not a confidence interval for the mean.}
\end{tabular}
\end{center}
\end{table}

Nested LOOCV gave similar results (Table~\ref{tab:nested}). On $n=118$, elastic net, random forest, and gradient boosting reached AUCs of 0.34, 0.55, and 0.62, with permutation $p$ of 0.990, 0.248, and 0.109. For gradient boosting, inner tuning most often chose a learning rate of 0.05 (82 of 118 folds), 100 trees (85 of 118), and a tree depth of 2 (83 of 118); all three together were chosen in 34 of 118 folds. The below-chance elastic net AUC matches the known downward bias of LOOCV AUC when there is no signal \cite{b_parker,b_airola}. It should not be read as a reversed signal. Under repeated five-fold cross-validation, the same model gives 0.376, with each fold's AUC computed on many test samples instead of one. On C1 alone, elastic net LOOCV gave AUC 0.449 ($p=0.560$).

None of the six correction methods helped (Table~\ref{tab:correction}). Uncorrected data gave the best random forest AUC, and cohort-as-covariate gave the worst. With cohort as a covariate, the random forest split on the cohort indicators and predicted each cohort's majority class, which fails because the fraction of responders differs by cohort. Percentile normalization gave the best elastic net AUC but cut response $R^2$ by 81\%, which shows why a correction should not be judged by one number alone. After every correction, response $R^2$ stayed at or below its chance level.

\subsection{Cross-Cohort Transfer}

We tested nine train/test splits on C1 to C3 (three leave-one-cohort-out splits and six single-cohort-to-single-cohort splits), each with elastic net and random forest, for 18 results in total; C4 was not used here. None reached significance (best $p=0.338$), mean AUC was 0.395 for elastic net and 0.382 across both models, and 16 of 18 fell below 0.5. For two of the three held-out cohorts, training on one other cohort did better than training on all the others combined (Table~\ref{tab:transfer}).

Results that are consistently below chance are worth explaining, because pure noise would scatter around 0.5. There are two reasons. First, the 18 results are not independent, since each cohort is the test set in three splits, or six results. Second, below-chance transfer is what you would expect if the direction of a bacterium's association with response flips between cohorts. We checked whether one cohort caused the pattern and found it did not. Mean AUC across both models was 0.376, 0.384, and 0.386 when C1, C2, and C3 were the test cohort, with five, six, and five of six splits below chance. C2, the gastrointestinal cohort, was slightly more extreme, with every split below chance, which fits with it being the most biologically different. However, the difference is small. At the level of individual genera, sign flips were no more common than with shuffled labels (Section~\ref{sec:features}). We therefore interpret the transfer results as cohort-specific patterns that do not generalize, not as a consistent opposite biology.

\subsection{The Largest Single Cohort}
\label{sec:c4}

C4 ($n=165$) has no between-study differences, but it was recruited at five sites, and those sites behave just like cohorts. Recruiting site explained $R^2 = 0.094$ against a chance level of 0.024, an excess of 7.0 percentage points ($p=0.001$). Response explained $R^2 = 0.0066$ against a chance level of 0.0061, an excess of only 0.05 points ($p=0.277$; Fig.~\ref{fig:pcoa}C). Sites did not differ in spread (PERMDISP $F=2.28$, $p=0.120$), so the site effect reflects differences in where the sites' communities are centered, not in how variable they are. So the same pattern that appears when separate studies are pooled also appears inside a single study that recruited at multiple sites.

Five preprocessing variants gave elastic net AUCs of 0.56 to 0.65, with permutation $p$ of 0.065 to 0.184 (Table~\ref{tab:c4}). None were significant. The best one followed rarefaction to one million reads; with five uncorrected variants, a smallest $p$ of 0.065 is about what the best of five null results would give.

\subsection{Individual Genera}
\label{sec:features}

If a weak but real signal existed, individual genera should be linked to response in the same direction across cohorts, even if classification fails. They are not. Of 48 genera with nonzero elastic net coefficients in at least two per-cohort models, 64.6\% flipped sign between cohorts, compared with 58.8\% under shuffled labels ($p=0.213$).

\begin{figure}[!htbp]
\centerline{\includegraphics[width=\columnwidth]{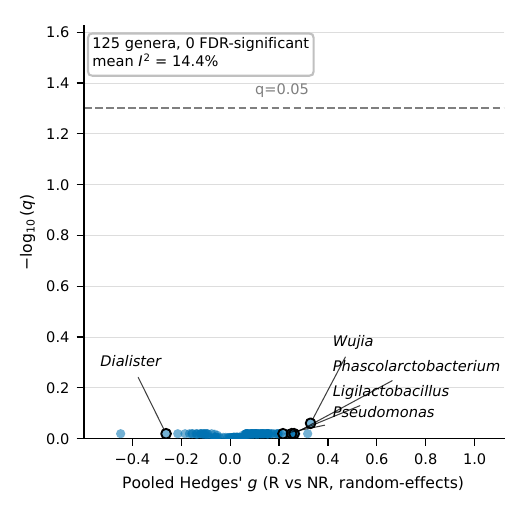}}
\caption{Random-effects meta-analysis of 125 genera present in at least two cohorts (Hedges' $g$ for responders versus non-responders on CLR abundance). No genus comes close to the FDR threshold (dashed line, $q=0.05$). The five with the smallest $q$ are labeled.}
\label{fig:volcano}
\end{figure}

The random-effects meta-analysis gives the same answer. Of 125 genera present in at least two cohorts, none reached FDR significance, and mean $I^2$ was 14.4\% (Fig.~\ref{fig:volcano}). The genus closest to significance had $q=0.87$. For a genus present in all four cohorts, the smallest pooled effect detectable with 80\% power is about $g=0.34$ at $\alpha=0.05$, and larger after FDR correction; for a genus found only in the three small cohorts it is about 0.53. So this analysis can rule out moderate or large shared effects, not small ones. The low $I^2$ is worth pointing out: the cohorts mostly agree with each other about these genera, and what they agree on is that the effects are close to zero.

\subsection{Tumor Genomic Replication}

\begin{figure}[!htbp]
\centerline{\includegraphics[width=\columnwidth]{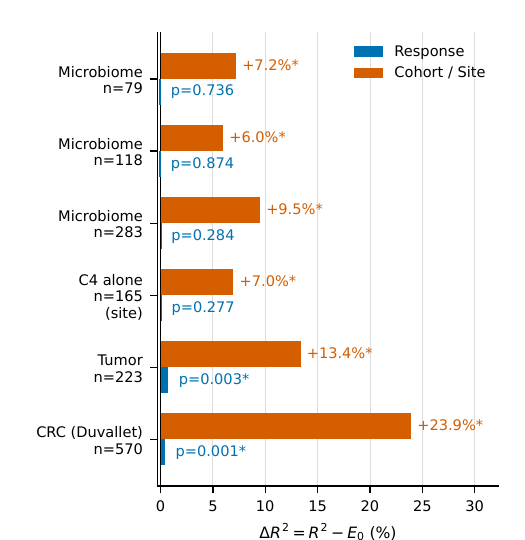}}
\caption{Chance-corrected excess variance explained by cohort (or recruiting site, for C4 alone) and by treatment response, across every dataset. Asterisks mark $p<0.05$. Response excess is significant only in the tumor and colorectal datasets, where the cohort effect is about eighteen and fifty times larger. In all four microbiome settings, response cannot be told apart from chance.}
\label{fig:summary}
\end{figure}

The tumor genomic analysis tests whether cohort dominance depends on problems specific to microbiome sequencing. It does not (Fig.~\ref{fig:summary}). Across three tumor studies ($n=223$), cohort membership explained $R^2 = 0.143$ against a chance level of 0.009, an excess of 13.4 percentage points. Unlike the microbiome data, tumor response was significant ($p=0.003$), with $R^2 = 0.012$ against a chance level of 0.0045, an excess of 0.75 points, and it stayed significant in the joint model. So a real but small response effect exists alongside a cohort effect about eighteen times larger. This is the same situation as the colorectal validation in Section~\ref{sec:tool}.

Our submitted version compared raw $\rho$ between the tumor data ($\rho=11.9$, $n=223$) and the three-cohort microbiome set ($\rho=11.3$, $n=118$). We no longer treat that match as evidence, because $\rho$ depends on $n$ and on whether any response effect exists; the microbiome value at $n=283$ is 27.3. The chance-corrected comparison is the meaningful one. It shows a cohort excess of several percentage points or more in every dataset, and a response excess that is zero in the microbiome cohorts and small where it exists.

Pooled tumor classification gave AUCs of 0.45 to 0.52, at or near chance. Only gradient boosting was permutation-tested (AUC 0.448, $p=0.745$). The largest single tumor cohort, Liu 2019 ($n=144$), reached elastic net AUC 0.513 ($p=0.290$). The most frequently mutated genes in Liu 2019 were TTN (63.9\% of patients) and MUC16 (59.7\%). These are very long genes that are often mutated by chance, so they mostly track overall mutation count. In a gradient boosting model fit to Liu 2019, the most important feature was RAC1, a known melanoma driver gene, and MAP2K1 and KIT, two other known drivers, were also in the top ten. The rest of the top ten were mostly large, frequently mutated genes such as DNAH5, OBSCN, and ADGRV1. When each study was held out and models were trained on the other two, models tested on Hugo reached AUC 0.58 to 0.62, while models tested on Riaz dropped to 0.34 to 0.45. The cohort effect also works differently in each data type. Tumor cohorts mainly differ in magnitude, with median tumor mutational burden of 4.7, 6.5, and 13.1 in Riaz, Liu, and Hugo, while microbiome cohorts differ in composition.

\subsection{Simulation}

\begin{figure*}[!t]
\centerline{\includegraphics[width=\textwidth]{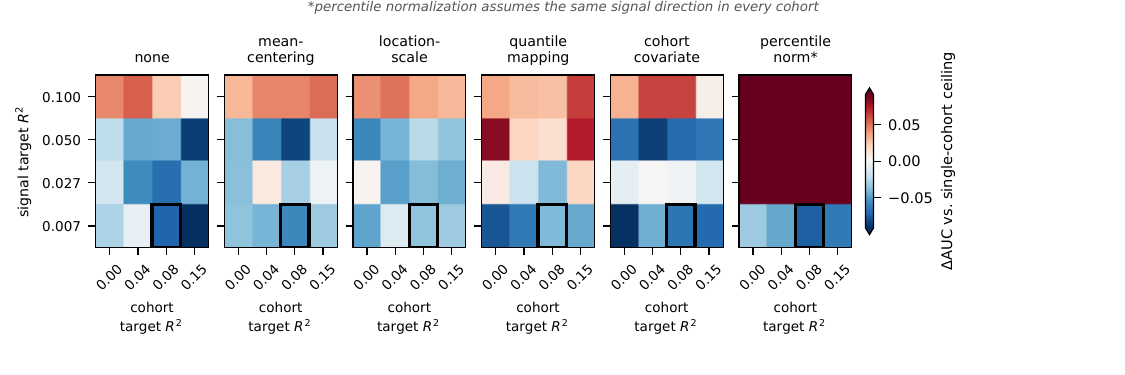}}
\caption{Calibrated simulation: change in AUC compared with the single-cohort, no-cohort-effect ceiling, for three cohorts of 40 samples, by signal size (rows) and cohort effect size (columns). Red cells mean correction beats the ceiling; blue means it falls short. The bottom row ($R^2_{\mathrm{t}}=0.007$) was calibrated to our pooled $n=118$ response $R^2$, which is at chance, so it contains essentially no signal. Black outline: the setting that matches our real data. The color scale is set by the first five methods; percentile normalization's large gains at higher signal levels exceed it and are clipped (arrow caps on the color bar). Those gains depend on the signal pointing the same way in every cohort, which the simulation assumes but our real data do not show.}
\label{fig:sim}
\end{figure*}

The simulation shows that our data sit outside the range where correction helps (Fig.~\ref{fig:sim}, Table~\ref{tab:sim}). With three cohorts of 40, pooled models reliably beat the single-cohort ceiling only at the largest simulated signal, $R^2_{\mathrm{t}} = 0.10$, and there they did so with or without correction. At $R^2_{\mathrm{t}} = 0.05$, results were mixed. Quantile mapping helped in several settings while the other methods mostly did not. However, the simulation assumes the signal points the same way in every cohort, and our cross-cohort results suggest it does not, so this advantage may not carry over to real data. At $R^2_{\mathrm{t}} = 0.027$, all methods were near or below the ceiling. At $R^2_{\mathrm{t}} = 0.007$, which matches our real data, every method fell below the ceiling, by 0.036 to 0.075 at a cohort effect of $R^2_{\mathrm{t}} = 0.08$. Percentile normalization showed very large gains at the higher signal levels, but these come from the same assumption that the signal points the same way in every cohort, and at the real-data setting it fell 0.075 below the ceiling like the others.

\subsection{Power}

\begin{figure}[!htbp]
\centerline{\includegraphics[width=\columnwidth]{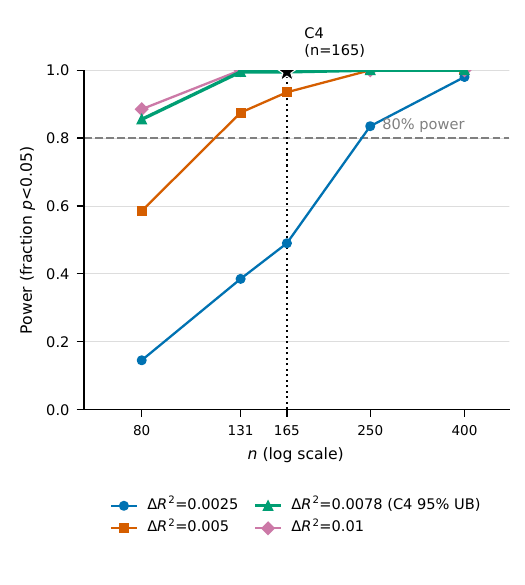}}
\caption{Simulated power of PERMANOVA ($\alpha=0.05$) to detect a hypothetical response effect $\Delta R^2$, by sample size, in C4's full genus space. Dashed line: 80\% power. Dotted line: C4's actual size. Star: power at C4's actual size for $\Delta R^2 = 0.0078$, the 95\% upper bound on C4's response effect (estimated in its 30 most common genera, where power at this point was 90\%).}
\label{fig:power}
\end{figure}

The power analysis asks a different question: was C4 simply too small to find a real effect? The answer is no (Table~\ref{tab:power}, Fig.~\ref{fig:power}). Bootstrapping C4's real data, using its 30 most common genera, gives a 95\% upper bound of $\Delta R^2 = 0.0078$ on its true response effect. In simulation in that same 30-genus space, an effect of that size was detected 90\% of the time at C4's size of $n=165$. In C4's full genus set, which the C4 PERMANOVA uses, power was even higher: an effect of that size was detected 99.5\% of the time at $n=165$, and the smallest effect detectable with 80\% power was about $\Delta R^2 = 0.0042$. C4 still found nothing ($p=0.277$). That is not what we would expect if a real effect near the top of its plausible range existed and the study was just underpowered. It is what we would expect if the true effect is close to the observed value, which is essentially zero.

\section{Discussion}

Our results make one argument. For overall community composition, response explains no more variance than chance at any cohort size, in any single cohort, at equal sequencing depth, and at every taxonomic level and zero-handling method, and adding cohort to the model does not change that. For classification, no model, correction method, or validation design finds a signal. For individual genera, none pass FDR correction, and their directions are no more consistent across cohorts than with shuffled labels. A weak but real signal would be expected to survive at least the last test, because consistent directions can show up even when classification fails. Since all three tests fail, the pooled data seem to contain no response signal that can be recovered, rather than a signal too subtle to model. Throughout, cohort membership explains several percentage points of variance beyond chance.

Response $R^2$ drops as cohorts are pooled because its chance level drops as $1/(n-1)$, and the ratio of cohort $R^2$ to response $R^2$ grows with $n$ for the same reason. The same issue applies to any study that compares PERMANOVA $R^2$ across sample sizes or across factors with different numbers of groups. The chance correction in \eqref{eq:null} is standard for ANOVA but rarely applied to PERMANOVA.

The tumor genomic analysis shows that cohort dominance is not just a microbiome problem. Microbiome sequencing has known weaknesses, such as DNA extraction bias, contamination from human DNA, and gaps in the classification database, but none of these apply to tumor mutation calls. Still, the tumor cohorts also show a cohort effect many times larger than the response effect. The two data types differ in one important way: the tumor data contain a small, real response signal, and the microbiome data do not. What they share is the study design: a few small cohorts, collected at different sites under different protocols, and combined afterward.

These cohort effects are not only technical. Our cohorts differ in tumor type (C2), recruiting sites (C4), country, and, because of the shared curation, the definition of response, as well as in sequencing platform. Two cohorts sequenced on the same instrument model (C2 and C4) still separate clearly, so using the same machine does not make cohorts interchangeable. The C4 result shows this most directly. A single published study, with one protocol and one sequencing platform, still shows a site effect of seven percentage points above chance, against a response excess that cannot be told apart from zero. This is why we call it a cohort effect: it is the variation between studies, and sometimes within them, that anyone pooling these datasets inherits.

The practical takeaway is different from the usual advice to collect more samples. Our power analysis shows that C4, at $n=165$, was already large enough to detect any effect within the range its own data allow. The problem in these cohorts is not that the sample is too small for the effect. Based on our evidence, the response effect in these four public immunotherapy gut microbiome cohorts is close to zero, and the variation that does exist is organized by site and study instead of by treatment response. This conclusion applies to these datasets, this response definition, and taxonomic features. It does not rule out an effect in carefully harmonized future cohorts, at finer resolution, or in functional data. We recommend that multi-cohort omics studies report chance-corrected cohort and response variance, including a joint model with within-cohort shuffling, before reporting prediction accuracy. We also recommend that future studies standardize sample collection, sequencing, and response definitions from the start.

\subsection*{Why Our Result Differs from Published Positive Reports}

Several studies report microbiome signatures of response that hold across cohorts. Limeta et al. \cite{b_limeta} found taxa such as \textit{Faecalibacterium} and B-vitamin pathways enriched in responders, with modest predictive performance in a validation cohort, and Olekhnovich et al. \cite{b_olekh} reported \textit{Faecalibacterium prausnitzii}, \textit{Bifidobacterium adolescentis}, and \textit{Eubacterium rectale} as consistent species-level markers. Other work agrees more with us: Lee et al. \cite{b_lee}, the source of C4, found that associations and prediction varied between their cohorts, and a meta-analysis of 15 melanoma cohorts found signatures that depend on treatment context rather than any single universal species \cite{b_zhang}.

These results are not necessarily in conflict with ours, for four reasons. First, we ask whether response organizes overall community composition, while most positive reports test individual taxa or pathways; a few species with real effects can exist without shifting whole-community distances, and our genus meta-analysis can only rule out effects of about $g=0.34$ or larger. Second, our labels count stable disease as response, which is closer to clinical benefit than to objective response. Third, several of those studies use species-level or functional features, while ours are mostly genus-level. Fourth, our analysis corrects $R^2$ for chance and fits every preprocessing step inside cross-validation folds; our own early leakage result (AUC 0.995) shows how much a pooled analysis can gain from skipping the second step.

\subsection*{Corrections to Our Initial Analysis}

Three results in our submitted version were wrong, and we correct them here. First, we described the drop in response $R^2$ from 2.67 to 0.39\% as a 6.9-fold loss of biological signal caused by pooling. That drop is the chance level falling as $1/(n-1)$. Second, our first meta-analysis reported four FDR-significant genera and a mean $I^2$ of 99.5\%. It estimated each cohort's standard errors from how much elastic net coefficients varied across LOOCV folds, divided by $\sqrt{n}$. Those folds share $n-1$ of $n$ samples, so they are not independent, and dividing by $\sqrt{n}$ shrinks the errors a second time; errors that are too small inflate $I^2$ and create false positives. Third, we estimated that about 80 same-protocol patients would give 80\% power. That estimate was based on the observed $n=39$ effect, which is at chance, so it did not measure anything real.

\subsection*{Limitations}

Cohort C2 has a different tumor type, and all labels follow one curation's broader response definition, which we did not re-check against the original papers. Both add to the cohort effect along with technical differences. The capped 8~GB Kraken2 database is less sensitive than the full version. We processed four of the eleven studies in the label curation. The other seven have public reads but include combination therapies, chemotherapy, non-melanoma cancers, and possibly overlapping patients, and remain to be analyzed. We could not test a strict objective-response definition (complete or partial response versus progression), because per-patient labels were behind paywalls or could not be linked to sequencing runs. Functional pathway profiling was too computationally expensive at full scale on our hardware, so our analysis is taxonomic only, and we cannot rule out a functional signal that does not show up in taxonomy. The permutation tests for repeated cross-validation re-ran feature selection and correction in every fold but reused fixed hyperparameters instead of re-tuning them, which may underestimate how much results vary under the null. Simulated power depended on the feature space. We estimated C4's upper bound in its 30 most common genera, so our main power figure (90\%) comes from that same space; power in the full genus set, which the C4 PERMANOVA uses, was higher (99.5\%), and an earlier estimate using C1's 30 most common genera was lower. The number of features explains most of this difference and the choice of cohort explains the rest. The simulations also use a Gaussian model in CLR space, which does not capture every property of real count data. The joint model is our own implementation rather than an established package, although it reproduces the single-factor results as expected. Finally, our conclusions apply to public anti-PD-1 and anti-PD-L1 cohorts, not to studies designed with harmonized protocols from the start.

\section{Conclusion}

We pooled four public gut microbiome cohorts of immunotherapy patients, from 39 to 283 samples. At every step, treatment response explained no more variance than chance, while cohort membership explained 6.0 to 9.5 percentage points beyond chance. This null held for community composition, classification, and individual genera; within each cohort alone; across model types, correction methods, taxonomic levels, zero-handling methods, sequencing depth, and cross-validation designs; and in the largest single cohort, where recruiting site alone produced a cohort-like effect of seven percentage points. That cohort had at least 90\% simulated power to detect an effect at its own 95\% upper bound, so its null result points to an effect near zero, not to a study that was too small. In tumor genomic data, a small real response effect was outweighed by a cohort effect about eighteen times larger. For these four cohorts, building a pooled predictor is therefore not a workable approach, and PERMANOVA $R^2$ values should always be compared with their chance level. We release \bd{} so that this check takes minutes instead of months.

\section*{Acknowledgment}

Analysis scripts were written by the author, with Anthropic's Claude Code used to help debug them and improve their readability. During revision, Claude Code was also used to implement some additional analyses the author specified and to generate some figures from the author's result files; the author reviewed all code and outputs. Where standard methods have known failure modes on compositional data, Anthropic's Claude was consulted to identify candidate approaches from the literature: Dirichlet resampling of compositional weights for the bootstrap intervals, and Ledoit-Wolf shrinkage for the simulation covariance estimate. The author verified each empirically before adopting it, confirming for example that naive resampling produced the degenerate zero-distance pairs described in Section~\ref{sec:varpart}. Claude was also used to draft and edit manuscript prose, and flagged a few factual inconsistencies between draft text and source results. The author is responsible for all analytical decisions, verification of method appropriateness, and interpretation of results presented here. Several revisions in this version were made in response to comments from the IEEE BIBM 2026 reviewers.

\section*{Appendix: Supplementary Tables}

Tables \ref{tab:percohort}--\ref{tab:power} give the full numbers behind the results in Section~\ref{sec:results}.

\begin{table}[!htbp]
\caption{Response PERMANOVA Within Each Cohort (Genus Level, Aitchison Distance, 999 Permutations)}
\label{tab:percohort}
\begin{center}
\begin{tabular}{lccccc}
\toprule
Cohort & $n$ & $R^2$ & $E_0[R^2]$ & $\Delta R^2$ & $p$ \\
\midrule
C1 & 39  & 0.0267 & 0.0263 & $+0.0004$ & 0.419 \\
C2 & 40  & 0.0279 & 0.0256 & $+0.0023$ & 0.275 \\
C3 & 39  & 0.0238 & 0.0263 & $-0.0025$ & 0.635 \\
C4 & 165 & 0.0066 & 0.0061 & $+0.0005$ & 0.277 \\
\bottomrule
\end{tabular}
\end{center}
\end{table}

\begin{table}[!htbp]
\caption{Range of $\Delta R^2$ Across Three Zero-Handling Methods ($\epsilon=0.5$, $\epsilon=1$, Multiplicative Replacement; All Cohort $p=0.001$)}
\label{tab:zero}
\begin{center}
\begin{tabular}{llcc}
\toprule
$n$ & Factor & min $\Delta R^2$ & max $\Delta R^2$ \\
\midrule
39  & Response & $-0.0061$ & $-0.0019$ \\
79  & Response & $-0.0028$ & $-0.0025$ \\
79  & Cohort   & $+0.0856$ & $+0.0887$ \\
118 & Response & $-0.0015$ & $+0.0010$ \\
118 & Cohort   & $+0.0772$ & $+0.1176$ \\
283 & Response & $-0.0009$ & $+0.0003$ \\
283 & Cohort   & $+0.1070$ & $+0.1844$ \\
\bottomrule
\end{tabular}
\end{center}
\end{table}

\begin{table}[!htbp]
\caption{Taxonomic Level ($n=118$; $E_0$: Response 0.0085, Cohort 0.0171)}
\label{tab:taxa}
\begin{center}
\begin{tabular}{lrcccc}
\toprule
Level & Features & Resp. $R^2$ & Resp. $p$ & Cohort $R^2$ & Cohort $\Delta R^2$ \\
\midrule
Phylum  & 65    & 0.0066 & 0.629 & 0.0569 & $+0.0398$ \\
Genus   & 2,813 & 0.0068 & 0.874 & 0.0768 & $+0.0597$ \\
Species & 8,051 & 0.0080 & 0.595 & 0.0740 & $+0.0569$ \\
\bottomrule
\end{tabular}
\end{center}
\end{table}

\begin{table}[!htbp]
\caption{Nested LOOCV Classification (Secondary Protocol)}
\label{tab:nested}
\begin{center}
\begin{tabular}{llcccc}
\toprule
$n$ & Model & AUC & Sens. & Spec. & Perm. $p$ \\
\midrule
118 & Elastic net       & 0.338 & 0.375 & 0.389 & 0.990 \\
118 & Random forest     & 0.548 & 0.703 & 0.370 & 0.248 \\
118 & Gradient boosting & 0.617 & 0.641 & 0.556 & 0.109 \\
283 & Elastic net       & 0.517 & 0.453 & 0.497 & 0.267 \\
283 & Random forest     & 0.541 & 0.391 & 0.677 & 0.168 \\
283 & Gradient boosting & 0.524 & 0.414 & 0.658 & 0.248 \\
\bottomrule
\end{tabular}
\end{center}
\end{table}

\begin{table}[!htbp]
\caption{Batch Correction Comparison ($n=118$, LOOCV, Leak-Free; a Separate Run from Table~\ref{tab:nested}, So the ``None'' Row Differs Slightly)}
\label{tab:correction}
\begin{center}
\begin{tabular}{lccc}
\toprule
Method & Resp. $R^2$ & EN AUC & RF AUC \\
\midrule
None                  & 0.0068 & 0.351 & 0.542 \\
Mean centering        & 0.0060 & 0.329 & 0.532 \\
Location-scale        & 0.0060 & 0.363 & 0.513 \\
Quantile mapping      & 0.0057 & 0.347 & 0.419 \\
Percentile norm.      & 0.0013 & 0.419 & 0.465 \\
Cohort covariate      & 0.0068 & 0.354 & 0.344 \\
\bottomrule
\multicolumn{4}{l}{Chance level of response $R^2$ at $n=118$: 0.0085.}
\end{tabular}
\end{center}
\end{table}

\begin{table}[!htbp]
\caption{Cross-Cohort Transfer (C1--C3; Elastic Net Only, Random Forest Results in Text)}
\label{tab:transfer}
\begin{center}
\begin{tabular}{lccc}
\toprule
Test cohort & Best single-source AUC & Source & Pooled AUC \\
\midrule
C1 & 0.550 & C2 & 0.308 \\
C2 & 0.400 & C1 & 0.411 \\
C3 & 0.425 & C2 & 0.367 \\
\bottomrule
\end{tabular}
\end{center}
\end{table}

\begin{table}[!htbp]
\caption{C4 Sensitivity Analysis (Elastic Net, LOOCV, 200 Permutations)}
\label{tab:c4}
\begin{center}
\begin{tabular}{lccc}
\toprule
Variant & $n$ & AUC & Perm. $p$ \\
\midrule
Baseline (prevalence 10\%, top 100)  & 165 & 0.570 & 0.119 \\
Prevalence 20\%, top 100             & 165 & 0.593 & 0.080 \\
Prevalence 10\%, top 50              & 165 & 0.562 & 0.184 \\
Rarefied to 1M reads                 & 165 & 0.651 & 0.065 \\
Three largest sites only             & 135 & 0.563 & 0.119 \\
\bottomrule
\end{tabular}
\end{center}
\end{table}

\begin{table}[!htbp]
\caption{Simulated Change in AUC vs. Single-Cohort Ceiling (3 Cohorts, 40 per Cohort, Cohort $R^2_{\mathrm{t}} = 0.08$)}
\label{tab:sim}
\begin{center}
\begin{tabular}{lcc}
\toprule
Method & Signal $R^2_{\mathrm{t}}=0.007$ (real data) & Signal $R^2_{\mathrm{t}}=0.027$ \\
\midrule
No correction     & $-0.072$ & $-0.069$ \\
Mean centering    & $-0.059$ & $-0.030$ \\
Location-scale    & $-0.036$ & $-0.040$ \\
Quantile mapping  & $-0.041$ & $-0.040$ \\
Percentile norm.$^{\mathrm{a}}$  & $-0.075$ & -- \\
Cohort covariate  & $-0.066$ & $-0.003$ \\
\bottomrule
\multicolumn{3}{l}{$^{\mathrm{a}}$Reported only at the real-data setting; all cells are in Fig.~\ref{fig:sim}.}
\end{tabular}
\end{center}
\end{table}

\begin{table}[!htbp]
\caption{Simulated Power in C4's Full Genus Set ($\alpha=0.05$, 200 Replicates per Cell; $\Delta R^2=0.0078$ Is C4's Bootstrap 95\% Upper Bound)}
\label{tab:power}
\begin{center}
\begin{tabular}{lcccc}
\toprule
$n$ & $\Delta R^2 = 0.0025$ & $0.005$ & $0.0078$ & $0.01$ \\
\midrule
80  & 0.145 & 0.585 & 0.855 & 0.885 \\
131 & 0.385 & 0.875 & 0.995 & 1.000 \\
165 & 0.490 & 0.935 & 0.995 & 1.000 \\
250 & 0.835 & 1.000 & 1.000 & 1.000 \\
400 & 0.980 & 1.000 & 1.000 & 1.000 \\
\bottomrule
\end{tabular}
\end{center}
\end{table}

\FloatBarrier

\end{document}